# Predictions on the alpha decay chains of superheavy nuclei with Z =121 within the range 290 ≤ A ≤ 339

K. P. Santhosh* and C. Nithya

*School of Pure and Applied Physics, Kannur University, Swami Anandatheertha Campus, Payyanur 670327, Kerala, India*

**Abstract**

A systematic study on the alpha decay half lives of various isotopes of superheavy element Z = 121 within the range 290 ≤ A ≤ 339 is presented for the first time using Coulomb and proximity potential model for deformed nuclei (CPPMDN). The calculated α decay half lives of the isotopes within our formalism match well with the values computed using Viola-Seaborg systematic, Universal curve of Poenaru et al., and the analytical formula of Royer. In our study by comparing the α decay half lives with the spontaneous fission half lives, we have predicted 2α chain from $^{309, 311, 312}$121, 3α chain from $^{310}$121 and 1α chain from $^{313, 314}$121. Clearly our study shows that the isotopes of superheavy element Z = 121 within the mass range 309 ≤ A ≤ 314 will survive fission and can be synthesized and detected in the laboratory via alpha decay. We hope that our predictions will provide a new guide to future experiments.

* email: drkpsanthosh@gmail.com

## 1. Introduction

Understanding the physical as well as the chemical properties of superheavy elements has now become one of the hardest challenges in nuclear science. The search for heavy and superheavy elements started in 1940s with the synthesis of neptunium (Z = 93) at Lawrence Berkeley Laboratory in Berkeley (USA) [1]. The process of synthesizing nuclei beyond uranium via nuclear reactions is a challenging one because the isotopes of elements with Z > 92 are too short lived to be detected. However theoretical results have led to the prediction of an "island of stability [2-6]" for superheavy elements, which should have half lives ranging from minutes to several years. The discovery of shell structure directs to the question "whether the shell effects can stabilize nuclei to exist in regions of macroscopic instability". Superheavy elements (SHE) usually refers to the elements far beyond uranium and the major aim of superheavy element research is the investigation of nuclear matter under large Coulomb force. The search for island of stability has led to the synthesis of elements up to Z =118, thereby confirming the existence of magic island which is the most demanding topic in heavy ion research.

Recently the isotopes of superheavy elements are produced via hot and cold fusion reactions. Heavy ion accelerators which deliver intense heavy ion beams are used for the complete fusion of heavy ions. The cold fusion reaction [7] opened up ways for the synthesis of SHN with Z = 107-112 [8], at GSI, Darmstadt and RIKEN, Japan whereas hot fusion reaction [9] has been successful in the synthesis of SHN with Z =113-118 [8] at JINR-FLNR, Dubna. An attempt for the production of the superheavy element with Z = 120 [10] through hot fusion reaction was done by Oganessian et al. in 2009.

Theoretical efforts on the predictions of various properties of SHN had acquired exciting progress in the last two decades. Many of these studies suggested favorable candidates for magic numbers, next to the presently known Z = 82 and N = 126. Phenomenological models such as finite range droplet model (FRDM) predicts shell closure at Z = 114 and N = 184 [11]. The Skyrme Hartree-Fock (SHF) method predicts magic numbers at Z = 120 or Z = 126 and N = 184 [12, 13]. The prediction of magic neutron number, N = 184, is same in most of the theoretical predictions but for protons both Z = 114, 120 and 126 are suggested as magic numbers. The difficulty involved in localizing the energies of the single-particle levels between Z = 114 and 126 is considered as the reason for this uncertainty in proton magic number. Thus it was realized that the borderlines of the magic island depends highly on the model that is used.

The best way of studying superheavy elements is through the characterization of their decay properties. Superheavy nuclei decay is mainly by the emission of α particles followed by subsequent spontaneous fission. Naturally, for the identification of new elements α decay has been an inevitable tool. Proper measurement of α decay properties provide valuable information on the structure of superheavy nuclei such as shell effects and stability, nuclear spins and parities, deformation, rotational properties, fission barrier etc. Presently several theoretical approaches which come under macro-micro method like the cluster model [14], fission model [15], the density-dependent M3Y (DDM3Y) effective model [16], the generalized liquid drop model (GLDM) [17] etc and self consistent theories like the relativistic mean field theory [18], Skyrme-Hartree-Fock mean field model [19] etc are used to explain the α decay from heavy and superheavy nuclei. By using 20 mass models and 18 empirical formulas, systematic calculations on the α decay energies ($Q_\alpha$) and alpha decay half lives of superheavy nuclei with Z ≥ 100 are performed by Wang et al [20]. To reproduce the experimental $Q_\alpha$ values of SHN, the authors found that WS4 mass model [21] is the most accurate one. This study also shows that among 18 formulae used to calculate the α decay half lives, SemFIS2 formulae [22] is the best one to predict α decay half lives. In addition, for predicting the α decay half-lives of SHN, UNIV2 formula [22] with fewest parameters and the VSS [23, 24], SP [25, 26] and NRDX [27] formulae with fewer parameters works well.

The studies on the emission of clusters heavier than α particle, the heavy particle radioactivity (HPR) are important in superheavy region. Calculations for superheavy nuclei with Z = 104-124 done by Poenaru et al.,[28] revealed a trend toward shorter half lives and larger branching ratio relative to α decay for heavier SHs. Through this study the authors confirmed that it is possible to find regions in which HPR is stronger than α decay.

As a part of sailing towards the shores of the island of stability, here we present the studies on the α decay properties of 49 isotopes of the superheavy element with Z = 121 for the first time within the Coulomb and proximity potential model for deformed nuclei (CPPMDN) [29] which is an extension of Coulomb and proximity potential model (CPPM) [30], proposed by Santhosh et al. The well established CPPMDN has proved to be successful in describing the α decay properties of superheavy nuclei [31-34]. An overview of the work is as follows. The detailed description of Coulomb and proximity potential model for deformed nuclei (CPPMDN) is given in section 2. Section 3 presents results and discussions on the α decay properties and decay modes of the isotopes of superheavy nuclei with Z = 121. A brief summary of the results is provided in section 4.

**2. The Coulomb and proximity potential model for deformed nuclei (CPPMDN)**

In CPPMDN the interacting potential between two nuclei is taken as the sum of deformed Coulomb potential, deformed two term proximity potential and centrifugal potential, for both the touching configuration and for the separated fragments. For the pre-scission (overlap) region, simple power law interpolation has been used.

The interacting potential barrier for two spherical nuclei is given by

$$V = \frac{Z_1 Z_2 e^2}{r} + V_p(z) + \frac{\hbar^2 \ell(\ell+1)}{2\mu r^2}, \quad \text{for } z > 0 \quad (1)$$

Here $Z_1$ and $Z_2$ are the atomic numbers of the daughter and emitted cluster, 'r' is the distance between fragment centres, 'z' is the distance between the near surfaces of the fragments, $\ell$ represents the angular momentum and $\mu$ the reduced mass. $V_P$ is the proximity potential given by Blocki et al., [35, 36] as

$$V_p(z) = 4\pi\gamma b \left[\frac{C_1 C_2}{(C_1 + C_2)}\right] \Phi\left(\frac{z}{b}\right) \quad (2)$$

with the nuclear surface tension coefficient,

$$\gamma = 0.9517 \left[1 - 1.7826 (N - Z)^2 / A^2\right] \text{ MeV/fm}^2 \quad (3)$$

Here N, Z and A represent the neutron, proton and mass number of the parent nuclei. $\Phi$ represents the universal proximity potential [36] given as

$$\Phi(\varepsilon) = -4.41 e^{-\varepsilon/0.7176}, \text{ for } \varepsilon > 1.9475 \quad (4)$$

$$\Phi(\varepsilon) = -1.7817 + 0.9270\varepsilon + 0.0169\varepsilon^2 - 0.05148\varepsilon^3, \text{ for } 0 \leq \varepsilon \leq 1.9475 \quad (5)$$

with $\varepsilon = z/b$, where $b \approx 1$ fm is the width (diffuseness) of the nuclear surface. The Süsmann central radii $C_i$ of the fragments are related to the sharp radii $R_i$ as

$$C_i = R_i - \left(\frac{b^2}{R_i}\right) \text{ fm} \quad (6)$$

For $R_i$, we use semi-empirical formula in terms of mass number $A_i$ as [35]

$$R_i = 1.28 A_i^{1/3} - 0.76 + 0.8 A_i^{-1/3} \text{ fm} \quad (7)$$

The potential for the internal part (overlap region) of the barrier is given as,

$$V = a_0 (L - L_0)^n, \text{ for } z < 0 \quad (8)$$

where $L = z + 2C_1 + 2C_2$ fm and $L_0 = 2C$ fm, the diameter of the parent nuclei. The constants $a_0$ and n are determined by the smooth matching of the two potentials at the touching point.

The barrier penetrability P Using the one dimensional Wentzel-Kramers-Brillouin approximation, is given as

$$P = \exp\left\{-\frac{2}{\hbar}\int_a^b \sqrt{2\mu(V-Q)}dz\right\} \quad (9)$$

Here the mass parameter is replaced by $\mu = m A_1 A_2 / A$, where m is the nucleon mass and $A_1$, $A_2$ are the mass numbers of daughter and emitted cluster respectively. The turning points "a" and "b" are determined from the equation, $V(a) = V(b) = Q$, where Q is the energy resleased.

The half life time is given by,

$$T_{1/2} = \left(\frac{\ln 2}{\lambda}\right) = \left(\frac{\ln 2}{\nu P}\right) \quad (10)$$

Here $\lambda$ is the decay constant and assault frequency $\nu = \left(\frac{\omega}{2\pi}\right) = \left(\frac{2E_\nu}{h}\right)$. The empirical vibration energy $E_\nu$, is given as [15]

$$E_\nu = Q\left\{0.056 + 0.039\exp\left[\frac{(4-A_2)}{2.5}\right]\right\}, \text{ for } A_2 \geq 4 \quad (11)$$

The Coulomb interaction between the two deformed and oriented nuclei taken from Ref. [37] with higher multipole deformations included [38, 39] is given as,

$$V_C = \frac{Z_1 Z_2 e^2}{r} + 3 Z_1 Z_2 e^2 \sum_{\lambda, i=1,2} \frac{1}{2\lambda+1} \frac{R_{0i}^{\lambda}}{r^{\lambda+1}} Y_{\lambda}^{(0)}(\alpha_i) \left[ \beta_{\lambda i} + \frac{4}{7} \beta_{\lambda i}^2 Y_{\lambda}^{(0)}(\alpha_i) \delta_{\lambda,2} \right] \quad (12)$$

with

$$R_i(\alpha_i) = R_{0i} \left[ 1 + \sum_{\lambda} \beta_{\lambda i} Y_{\lambda}^{(0)}(\alpha_i) \right] \quad (13)$$

where $R_{0i} = 1.28 A_i^{1/3} - 0.76 + 0.8 A_i^{-1/3}$. Here $\alpha_i$ is the angle between the radius vector and symmetry axis of the $i^{th}$ nuclei (see Fig.1 of Ref [38]) and it is to be noted that the quadrupole interaction term proportional to $\beta_{21}\beta_{22}$, is neglected because of its short-range character.

The two-term proximity potential for interaction between a deformed and spherical nucleus is given by Baltz et al., [40] as

$$V_{P2}(R,\theta) = 2\pi \left[ \frac{R_1(\alpha) R_C}{R_1(\alpha) + R_C + S} \right]^{1/2} \left[ \frac{R_2(\alpha) R_C}{R_2(\alpha) + R_C + S} \right]^{1/2}$$

$$\times \left[ \left[ \varepsilon_0(S) + \frac{R_1(\alpha) + R_C}{2 R_1(\alpha) R_C} \varepsilon_1(S) \right] \left[ \varepsilon_0(S) + \frac{R_2(\alpha) + R_C}{2 R_2(\alpha) R_C} \varepsilon_1(S) \right] \right]^{1/2} \quad (14)$$

where $\theta$ is the angle between the symmetry axis of the deformed nuclei and the line joining the centers of the two interacting nuclei, and $\alpha$ corresponds to the angle between the radius vector and symmetry axis of the nuclei (see Fig. 5 of Ref [40]). $R_1(\alpha)$ and $R_2(\alpha)$ are the principal radii of curvature of the daughter nuclei, $R_C$ is the radius of the spherical cluster, $S$ is the distance between the surfaces along the straight line connecting the fragments, and $\varepsilon_0(S)$ and $\varepsilon_1(S)$ are the one dimensional slab-on-slab function.

## 3. Results and discussion

In this work we have selected 49 isotopes of superheavy element with Z = 121 for studying the alpha decay properties. Alpha decay and spontaneous fission (SF) are the dominant decay modes of superheavy nuclei. It is well known that the nuclei with alpha decay half lives smaller than half lives for SF will survive fission and thus can be detected in the laboratory through alpha decay. We carry out the half life calculations of the isotopes of superheavy nuclei with Z = 121, within the range 290 ≤ A ≤ 339 using CPPMDN [29] and the values are then compared with the half lives calculated by means of CPPM [30], Viola-seaborg semi-empirical (VSS) relationship [23], the universal curve of Poenaru et al., [41, 42] and the analytical formula of Royer [43]. The SF half lives of these nuclei are calculated by using the semi-empirical formula of Xu et al [44].

### 3.1 Alpha Decay Half lives

The alpha decay lifetimes mainly depends on the alpha decay energy. The $Q_\alpha$ released from ground state to ground state decay is calculated using the mass excess of the parent, daughter and alpha particle. The effect of atomic electrons on the energy of α particles is also taken into account while calculating the decay energy, and is given as,

$$Q = \Delta M_p - (\Delta M_\alpha + \Delta M_d) + k(Z_p^\varepsilon - Z_d^\varepsilon) \quad (15)$$

The mass excess of the parent, daughter and the α particle are represented by $\Delta M_p$, $\Delta M_d$ and $\Delta M_\alpha$. For most of the nuclei under study, the mass excess values were taken from mass table of Wang et al., [45] and for those nuclei whose mass excess are not available in Ref [45], the corresponding values were obtained from Moller et al., [46]. The electron screening effect on the energy of alpha particle is not included in the mass excess given in Ref [45]. So it is incorporated by adding the term $k(Z_p^\varepsilon - Z_d^\varepsilon)$ in Eqn (15). The term $kZ^\varepsilon$ is the total binding energy of Z electrons in the atom. Here k = 8.7 eV and ε = 2.517 for nuclei with Z ≥ 60 and k = 13.6 eV and ε = 2.408 for nuclei with Z < 60 [47, 48]. For calculating the alpha decay half lives several phenomenological formulae can be used. The models used in the present paper are the following:

### 3.1.1 The Viola-Seaborg Semi-empirical (VSS) Relationship

One of the most frequently used formulae for calculating the alpha decay half lives is the five parameter formula proposed by Viola and Seaborg. It is given by,

$$\log_{10}(T_{1/2}) = (aZ+b)Q^{-1/2} + cZ + d + h_{\log} \tag{16}$$

Here Z is the atomic number of the parent nucleus and a, b, c, d are adjustable parameters. The hindrance factor for nuclei with unpaired nucleons [23] is given by the quantity $h_{\log}$. Instead of using original set of constants given by Viola and Seaborg [23], more recent values determined by Sobiczewski et al., [24] has been used here. The constants are a = 1.66175, b = -8.5166, c = -0.20228, d = -33.9069 and

$$h_{\log} = \begin{cases} 0, & \text{for } Z = even \quad N = even \\ 0.772, & \text{for } Z = odd \quad N = even \\ 1.066, & \text{for } Z = even \quad N = odd \\ 1.114, & \text{for } Z = odd \quad N = odd \end{cases} \tag{17}$$

### 3.1.2 The Universal (UNIV) Curve of Poenaru et al.,

The Universal (UNIV) curve of Poenaru et al., [49-52] derived by extending a fission theory to larger asymmetry is one of the important relationship for calculating the decay half lives and is given by,

$$\log_{10} T(s) = -\log_{10} P_S - \log_{10} S + [\log_{10}(\ln 2) - \log_{10} \nu] \tag{18}$$

where T is the half life, ν, S and $P_S$ are three model dependent quantities. ν is the frequency of assaults on the barrier per second, S is the pre-formation probability of the cluster at the nuclear surface (equal to the probability of the internal part of the barrier in a fission theory [49, 50]), and $P_S$ is the quantum penetrability of the external potential barrier.

The penetrability of an external Coulomb barrier having the first turning point as the separation distance at the touching configuration $R_a = R = R_d + R_e$ and the second one defined by $e^2 Z_d Z_e / R_b = Q$ may be obtained analytically as,

$$-\log_{10} P_S = 0.22873(\mu_A Z_d Z_e R_b)^{1/2} \times [\arccos\sqrt{r} - \sqrt{r(1-r)}] \tag{19}$$

where $r = R_t / R_b$ fm, $R_t = 1.2249(A_d^{1/3} + A_e^{1/3})$ fm and $R_b = 1.43998 Z_d Z_e / Q$ fm.

The decimal logarithm of the pre-formation factor is given as,

$$\log_{10} S = -0.598(A_e - 1) \tag{20}$$

and the additive constant for even-even nuclei is,

$$c_{ee} = [-\log_{10} \nu + \log_{10}(\ln 2)] = -22.16917 \tag{21}$$

### 3.1.3 The Analytical Formula of Royer

An analytical formulae for α decay half lives have been developed by Royer [43] by applying a fitting procedure on α emitters, and is given by,

$$\log_{10}[T_{1/2}(s)] = a + bA^{1/6}\sqrt{Z} + \frac{cZ}{\sqrt{Q_\alpha}} \quad (22)$$

where A and Z represent the mass and charge number of the parent nuclei and $Q_\alpha$ represents the energy released during the reaction. The constants a, b and c are,

$$\left.\begin{array}{llll} a=-25.31, & b=-1.1629, & c=1.5864, & \text{for } Z=even \quad N=even \\ a=-26.65, & b=-1.0859, & c=1.5848, & \text{for } Z=even \quad N=odd \\ a=-25.68, & b=-1.1423, & c=1.5920, & \text{for } Z=odd \quad N=even \\ a=-29.48, & b=-1.1130, & c=1.6971, & \text{for } Z=odd \quad N=odd \end{array}\right\} \quad (23)$$

In the present work we have only used the analytical formulae for odd-even and odd-odd nuclei for calculating the alpha decay half lives.

### 3.2 Spontaneous fission half lives

Superheavy nuclei prominently decay through alpha emission followed by spontaneous fission. In the present work, we have used the semi-empirical relation given by Xu et al., [44] for calculating the SF half lives. The authors found that the calculated values of SF half lives using the semi-empirical formula matches well with the experimental results and also the logarithm of average deviations of 45 SF nuclei calculated using the formula is found to be 0.98 and this level of agreement is satisfactory because the SF is much more complex than other decay modes. We would like to mention that we have done a theoretical comparison [53] of calculated values of spontaneous fission half lives using different models. A large disagreement from model to model may be seen while taking the average deviation of spontaneous fission half lives calculated using different models [44, 54, 55].

#### 3.2.1 Semi-empirical Formula of Xu et al.,

The semi-empirical formula of Xu et al., [44] for calculating the SF half lives, is given by,

$$T_{1/2} = \exp\left\{2\pi\left[C_0 + C_1 A + C_2 Z^2 + C_3 Z^4 + C_4 (N-Z)^2 - \left(0.13323 \frac{Z^2}{A^{1/3}} - 11.64\right)\right]\right\} \quad (24)$$

The constants are $C_0$ = -195.09227, $C_1$ = 3.10156, $C_2$ = -0.04386, $C_3$ = 1.4030 x $10^{-6}$ and $C_4$ = -0.03199. The equation was originally made to fit the SF half lives of even-even nuclei. Since we have considered only the odd mass (i.e odd-even and odd-odd) nuclei in the present work, we have taken the average of SF half lives of the corresponding neighboring even-even nuclei. $T_{sf}^{av}$ of two neighboring even-even nuclei has been taken in the case of odd-even nuclei and $T_{sf}^{av}$ of four neighboring even-even nuclei has been taken while dealing with odd-odd nuclei.

### 3.3 Proton separation energy

The one proton and two proton separation energies [56] of all isotopes under study were evaluated to identify the proton emitters using the following relations

$$S(p) = -\Delta M(A,Z) + \Delta M(A-1, Z-1) + \Delta M_H = -Q(\gamma, p) \quad (25)$$
$$S(2p) = -\Delta M(A,Z) + \Delta M(A-2, Z-2) + 2\Delta M_H = -Q(\gamma, 2p) \quad (26)$$

where the terms $S(p)$ and $S(2p)$ are the one-proton separation energy and the two-proton separation energy of the nuclei, $\Delta M(A,Z)$, $\Delta M_H$, $\Delta M(A-1, Z-1)$ and $\Delta M(A-2, Z-2)$ represents the mass excess of the parent nuclei, the mass excess of the proton, the mass excess of the daughter nuclei produced during the one-proton radioactivity and the mass excess of the daughter nuclei produced during the two-proton radioactivity respectively. $Q(\gamma,$

*p*) and $Q(\gamma, 2p)$ represents respectively the $Q$ values for the one-proton radioactivity and two-proton radioactivity.

In the present work the mode of decay of 49 isotopes of Z = 121 within the range $290 \leq A \leq 339$ has been studied by evaluating the proton decay, alpha decay half lives and spontaneous fission (SF) half lives.

To know the behavior of $^{290-339}$121 SHN against proton decay, the proton separation energies of these isotopes are evaluated using Eqns (25) and (26). It was seen that the one-proton separation energy $S(p)$ is negative for the isotopes $^{290-303}$121 and the two-proton separation energy $S(2p)$ is negative for $^{290-299}$121. Thus it is clear that those isotopes of Z = 121 within the range $290 \leq A \leq 303$ may easily decay through proton emission.

The mode of decay of isotopes within the range $304 \leq A \leq 339$ has been studied by comparing the alpha decay half lives with the spontaneous fission half lives. Those isotopes with alpha decay half lives shorter than spontaneous fission half lives will survive fission and hence decay through alpha emission. In our study, using CPPMDN, we could observe 3α chains from $^{304}$121, 2α chains from $^{305}$121 and 3α chains from $^{306-308}$121. Since the alpha half lives of all these isotopes are less than micro second range (in the case of $^{304}$121, $T_{1/2}^{\alpha}$ = 1.161x10$^{-7}$ s, for $^{305}$121 $T_{1/2}^{\alpha}$ = 1.072x10$^{-7}$ s, for $^{306}$121 $T_{1/2}^{\alpha}$ =5.610x10$^{-9}$ s, for $^{307}$121 $T_{1/2}^{\alpha}$ = 4.779x10$^{-9}$ s, for $^{308}$121 $T_{1/2}^{\alpha}$ = 5.271x10$^{-7}$ s), these isotopes cannot be synthesized or detected in laboratories. Now, by comparing the alpha decay half lives with the spontaneous fission half lives, 2α chains can be seen from the isotopes $^{309}$121, 3α chains from $^{310}$121, 2α chains form $^{311,312}$121 and 1α chain from $^{313,314}$121. Since these isotopes show α decay followed by spontaneous fission and since their half lives are in measurable range, we can predict these isotopes to be synthesized and detected via alpha decay in laboratories. It is seen that the isotopes within the range $315 \leq A \leq 339$ will not survive fission and hence decay through spontaneous fission.

Thus from the entire study, it is evident that the isotopes of Z = 121 within the range $309 \leq A \leq 314$, shows α chain followed by SF with half lives in measurable range, are predicted to be synthesized and detected in laboratories. These predictions are shown in figures 1-6, which gives the plot of $\log_{10}(T_{1/2})$ versus mass number of the isotopes in the alpha decay chain. Alpha decay half lives evaluated using VSS, UNIV and Royer are shown in figures for a theoretical comparison with half lives calculated using our model. It is seen that the half lives calculated using other theoretical models matches well with our calculations. The SF half lives evaluated using the semi-empirical formula of Xu et al., are also shown, for comparing them with the alpha decay half lives and hence to predict the decay modes.

Table 1 gives the comparison of the alpha decay half lives with the SF half lives for the predicted isotopes $^{309-314}$121. The predictions on the mode of decay of these isotopes within CPPMDN are also given. The isotopes under study and the corresponding decay products are given in column 1. Coulmn 2 gives the theoretical Q values for these isotopes. In column 3, the SF half lives of the corresponding isotopes evaluated using the phenomenological formula of Xu et al., has been given. Column 4 presents the alpha decay half lives of these isotopes calculated using CPPMDN formalism. For calculating the alpha half lives within CPPMDN, the quadrupole ($\beta_2$) and hexadecapole ($\beta_4$) deformation values of the parent and daughter nuclei have been used, which are taken from Ref [46]. The calculations on the α half lives within CPPM is given in column 5. The nucleus-nucleus interaction potential in CPPMDN is calculated using equation (14) while equation (2) is used to calculate the potential in CPPM (spherical case). The alpha decay half lives calculated within VSS, UNIV and analytical formula of Royer are given in column 6, 7 and 8 respectively. The decay modes of isotopes under study are depicted in column 9.

The comparison of alpha decay half lives and spontaneous fission half lives for the isotopes of Z = 121 is done for the first time and we hope that our study in which we predicted the decay modes of various isotopes of Z = 121 will be a guideline for the future experiments.

**4 Conclusions**

In the present paper we described the theoretical predictions on the alpha decay half lives and decay modes of 49 isotopes of the superheavy element with Z = 121 within the Coulomb and proximity potential model for deformed nuclei (CPPMDN). The present values are then compared with the values calculated using other formalisms like Viola-Seaborg semi- empirical relationship, Universal Curve of Poenaru et al., and the analytical formula of Royer and found to be in good agreement. Through our study we inferred that the isotopes of Z = 121 within the range 309 ≤ A ≤ 314 will survive fission and thus can be synthesized and detected in laboratories. We have predicted 2α chain from $^{309}$121, 3α chain from $^{310}$121, 2α chain from $^{311, 312}$121 and 1α chain from $^{313, 314}$121. We hope that our predictions will open up new lines in experimental investigations.

**References**

[1] E. McMillan and P. H. Abelson, Phys. Rev. **57**, 1185 (1940).
[2] A. Sobiczewski, F. A. Gareev and B. N. Kalinkin, Phys. Lett. B **22**, 500 (1966).
[3] H. Meldner, Arkiv Fysik **36**, 593 (1967).
[4] W. D. Myers and W. J. Swiatecki, Arkiv Fysik **36**, 343 (1967).
[5] S. G. Nilsson, C. F. Tsang, A. Sobiczewski, Z. Szymański, S. Wycech, C. Gustafson, I. Lamm, P. Möller and B. Nilsson, Nucl. Phys. A **131,** 1 (1969).
[6] U. Mosel and W. Greiner, Z. Phys. **111**, 261 (1969).
[7] S. Hofmann and G. Munzenberg, Rev. Mod. Phys. **72**, 733 (2000).
[8] J. H. Hamilton, S. Hofmann and Yu. Ts. Oganessian, Annu. Rev. Nucl. Part. Sci. **63**, 383 (2013).
[9] Yu. Ts. Oganessian, J. Phys. G: Nucl. Part. Phys. **34**, R165 (2007).
[10] Yu. Ts. Oganessian, V. K. Utyonkov, Yu. V. Lobanov, F. Sh. Abdullin, A. N. Polyakov, R. N. Sagaidak, I. V. Shirokovsky, Yu. S. Tsyganov, A. A. Voinov, A. N. Mezentsev, V. G. Subbotin, A. M. Sukhov, K. Subotic, V. I. Zagrebaev, and S. N. Dmitriev, R. A. Henderson, K. J. Moody, J. M. Kenneally, J. H. Landrum, D. A. Shaughnessy, M. A. Stoyer, N. J. Stoyer and P. A. Wilk, Phys. Rev. C **79**, 024603 (2009).
[11] P. Moller and J. R. Nix, J. Phys. **G20**, 1681(1994).
[12] K. Rutz, M. Bender, T. Burvenich, T. Schilling, P. G. Reinhard, J. A. Maruhn, and W.Greiner, Phys. Rev. C **56,** 238(1997).
[13] S. Cwiok, J. Dobaczewski, P. H. Heenen, P. Magierski, and W. Nazarewicz, Nucl. Phys. A **611,** 211 (1996).
[14] B. Buck, A. C. Merchant and S. M. Perez, Phys. Rev. C **45**, 2247 (1992).
[15] D. N. Poenaru, M. Ivascu, A. Sandulescu and W. Greiner, Phys. Rev. C **32**, 572 (1985).
[16] D. N. Basu, Phys. Lett. B **566**, 90 (2003).
[17] H. F. Zhang and G. Royer, Phys. Rev. C **76,** 047304 (2007).
[18] M. M. Sharma, A. R. Farhan and G. Munzenberg, Phys. Rev. C **71,** 054310 (2005).

[19] J. C. Pei, F. R. Xu, Z. J. Lin and E. G. Zhao, Phys. Rev. C **76,** 044326 (2007).
[20] Y. Z. Wang, S J. Wang, Z. Y. Hou and J. Z. Gu, Phys. Rev. C **92**, 064301 (2015).
[21] N. Wang, M. Liu, X. Z. Wu and J. Meng, Phys. Lett. B **734**, 215 (2014).
[22] D. N. Poenaru, R. A. Gherghescu and N. Carjan, Euro Phys. Lett. **77**, 62001 (2007).
[23] V. E. Viola, Jr. and G. T. Seaborg, J. Inorg. Nucl. Chem. **28**, 741 (1966).
[24] A. Sobiczewski, Z. Patyk, and S. Cwiok, Phys. Lett. B 224, 1 (1989).
[25] A. Sobiczewski and A. Parkhomenko, Prog. Part. Nucl. Phys. **58**, 292 (2007).
[26] A. Parkhomenko and A. Sobiczewski, Acta Phys. Pol. B **36**, 3095 (2005).
[27] D. Ni, Z. Ren, T. Dong and C. Xu, Phys. Rev. C **78**, 044310 (2008).
[28] D. N. Poenaru, R. A. Gherghescu and W. Greiner, Phys. Rev. Lett. **107**, 062503 (2011).
[29] K. P. Santhosh, S. Sabina and G. J. Jayesh, Nucl. Phys. A **850**, 34 (2011).
[30] K. P. Santhosh and A. Joseph, Pramana. **62**, 957 (2004).
[31] K. P. Santhosh and B. Priyanka, Nucl. Phys. A **940**, 21 (2015).
[32] K. P. Santhosh, S. Indu and B. Priyanka, Nucl. Phys. A **935**, 28 (2015).
[33] K. P. Santhosh and B. Priyanka, Nucl. Phys. A **929**, 20 (2014).
[34] K. P. Santhosh and B. Priyanka, Phys. Rev. C **87**, 064611 (2013).
[35] J. Blocki, J. Randrup, W. J. Swiatecki and C. F. Tsang, Ann. Phys. (NY) **105**, 427 (1977).
[36] J. Blocki and W. J. Swiatecki, Ann. Phys. (NY) **132**, 53 (1981).
[37] C. Y. Wong, Phys. Rev. Lett. **31**, 766 (1973).
[38] N. Malhotra and R. K. Gupta, Phys. Rev. **C 31**, 1179 (1985).
[39] R. K. Gupta, M. Balasubramaniam, R. Kumar, N. Singh, M. Manhas and W. Greiner, J. Phys. G: Nucl. Part. Phys. **31**, 631 (2005).
[40] A. J. Baltz and B. F. Bayman, Phys. Rev. C **26**, 1969 (1982).
[41] D. N. Poenaru, R. A. Gherghescu and W. Greiner, Phys. Rev. C **83**, 014601 (2011).
[42] D. N. Poenaru, R. A. Gherghescu and W. Greiner, Phys. Rev. C **85**, 034615 (2012).
[43] G. Royer, J. Phys. G: Nucl. Part. Phys**. 26,** 1149 (2000).
[44] C. Xu, Z. Ren and Y. Guo, Phys. Rev. C **78**, 044329 (2008).
[45] M. Wang, G. Audi, A. H. Wapstra, F. G. Kondev, M. Mac Cormic, X. Xu and B. Pfeiffer, Chin. Phys. C **36**, 1603 (2012).
[46] P. Moller, A. J. Sierk, T. Ichikawa and H. Sagawa, At. Data Nucl. Data Tables **109**, 1 (2016).
[47] V. Yu. Denisov and A. A. Khudenko, Phys. Rev. C **79**, 054614 (2009).
[48] K. N. Huang, M. Aoyagi, M. H. Chen, B. Crasemann and H. Mark, At. Data Nucl. Data Tables **18**, 243 (1976).
[49] D. N. Poenaru and W. Greiner, J. Phys. G: Nucl. Part. Phys. **17**, S443 (1991).
[50] D. N. Poenaru and W. Greiner, Phys. Scr. **44**, 427 (1991).
[51] D. N. Poenaru, I. H. Plonski and W. Greiner, Phys. Rev. C **74**, 014312 (2006).
[52] D. N. Poenaru, I. H. Plonski, R. A. Gherghescu and W. Greiner, J. Phys. G: Nucl. Part. Phys. **32**, 1223 (2006).
[53] K. P. Santhosh, B. Priyanka and C. Nithya, Nucl. Phys. A (2016)
[54] M. Warda and J. L. Egido, Phys. Rev. C **86**, 014322 (2012).
[55] A. Staszczak, A. Baran and W. Nazarewicz, Phys. Rev. C **87**, 024320 (2013).
[56] S. Athanassopoulos, E. Mavrommatis, K. A. Gernoth, J. W. Clark, arXiv:nucl-th/0509075v1, 2005.

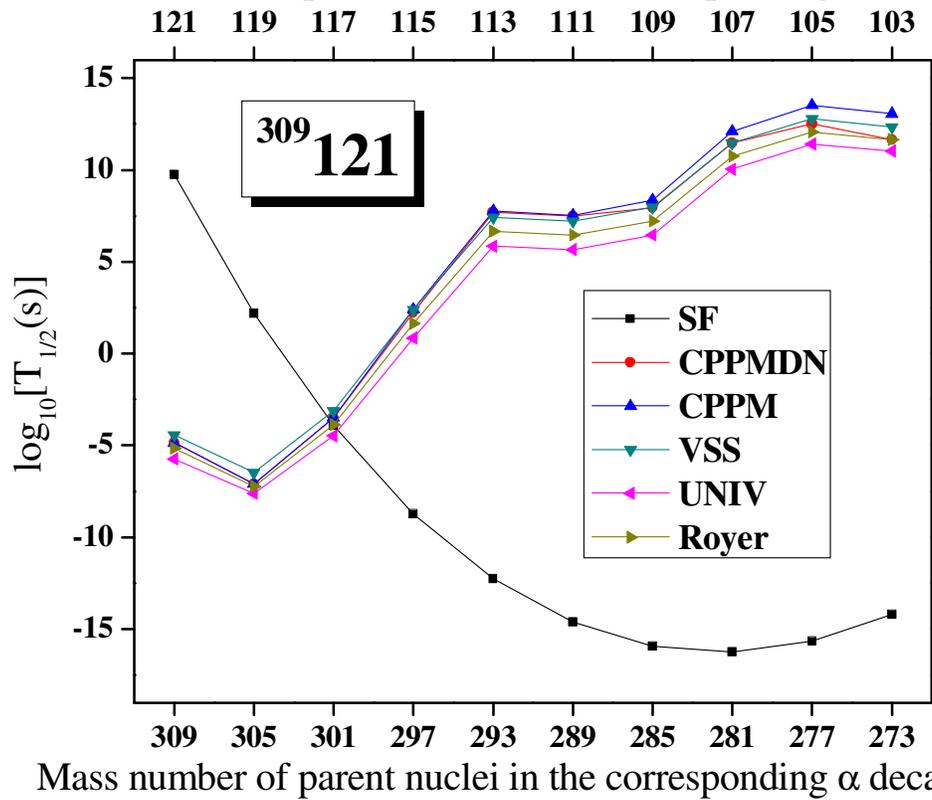

Fig 1: The comparison of the calculated alpha decay half-lives with the spontaneous fission half-lives for the isotope $^{309}121$ and its decay products.

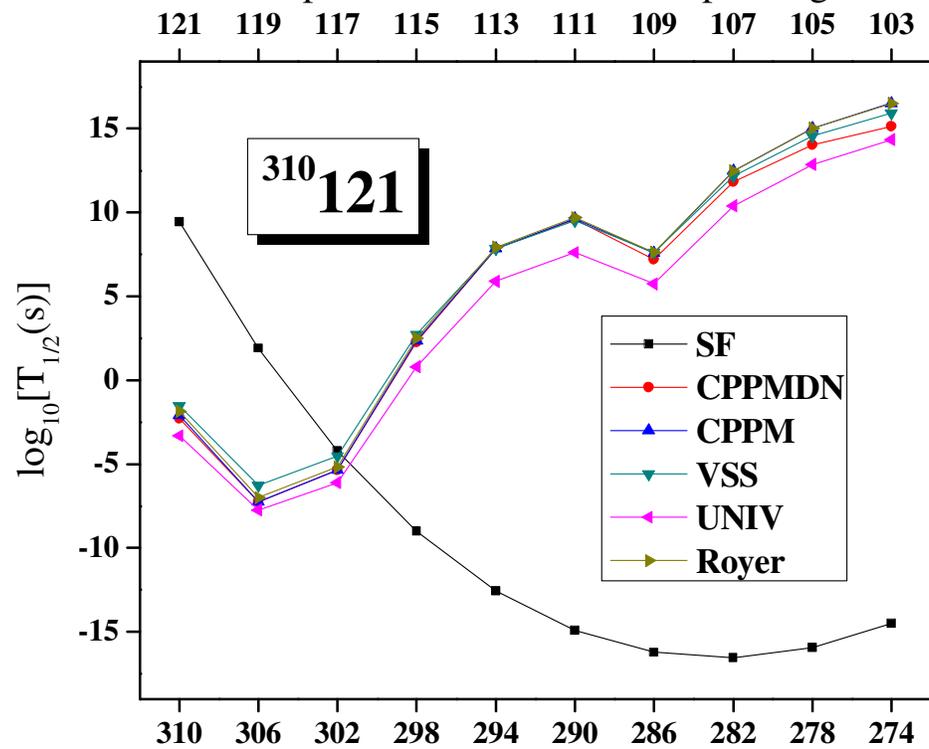

Fig 2: The comparison of the calculated alpha decay half-lives with the spontaneous fission half-lives for the isotope $^{310}$121 and its decay products.

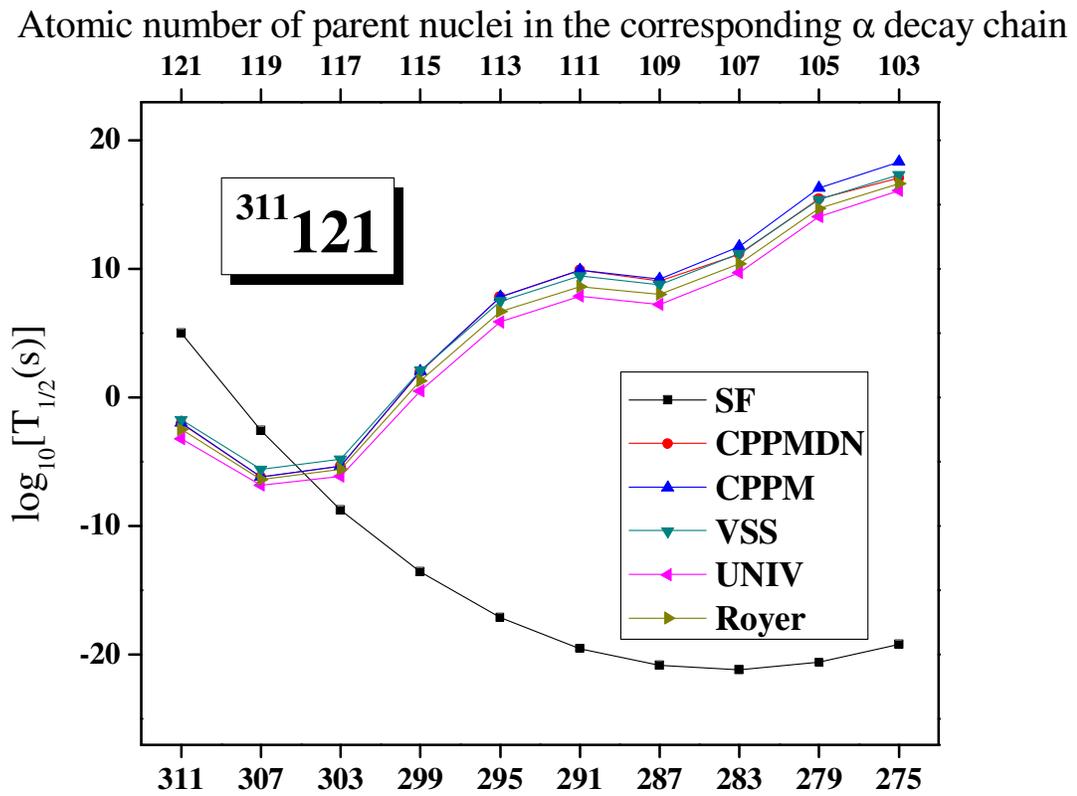

Fig 3: The comparison of the calculated alpha decay half-lives with the spontaneous fission half-lives for the isotope $^{311}121$ and its decay products.

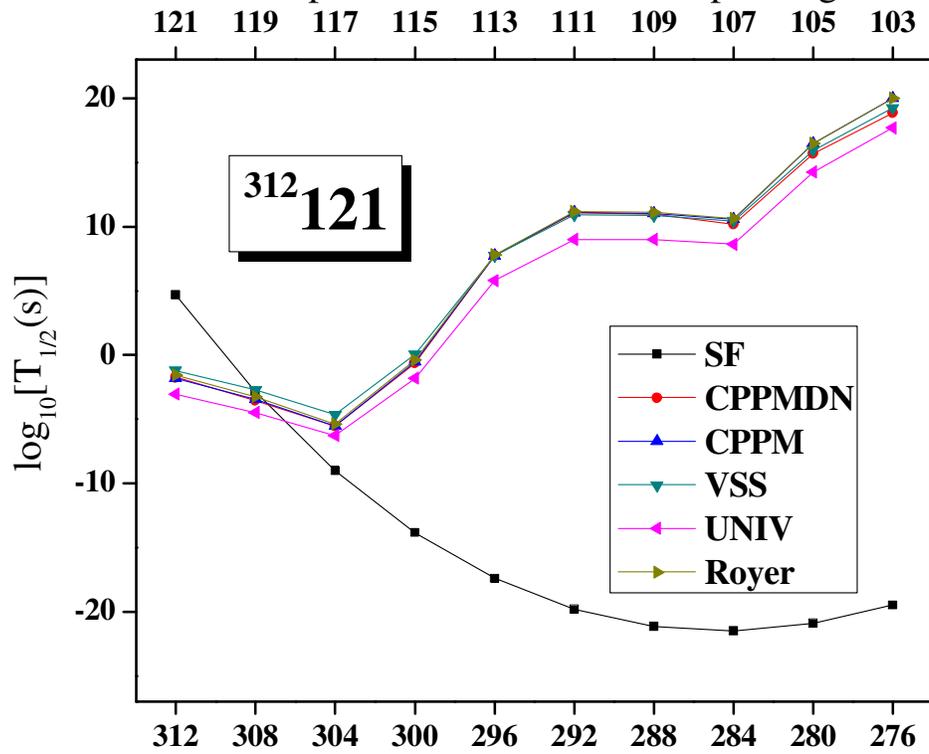

Fig 4: The comparison of the calculated alpha decay half-lives with the spontaneous fission half-lives for the isotope $^{312}121$ and its decay products.

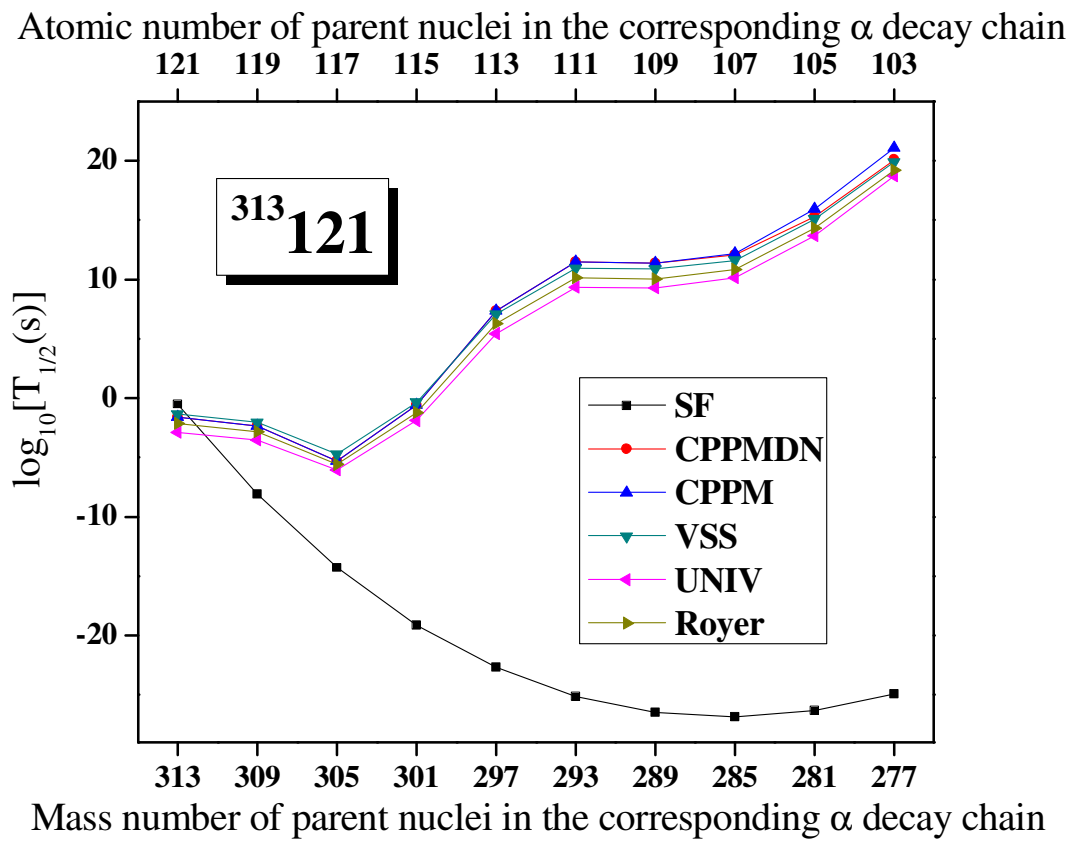

Fig 5: The comparison of the calculated alpha decay half-lives with the spontaneous fission half-lives for the isotope $^{313}121$ and its decay products.

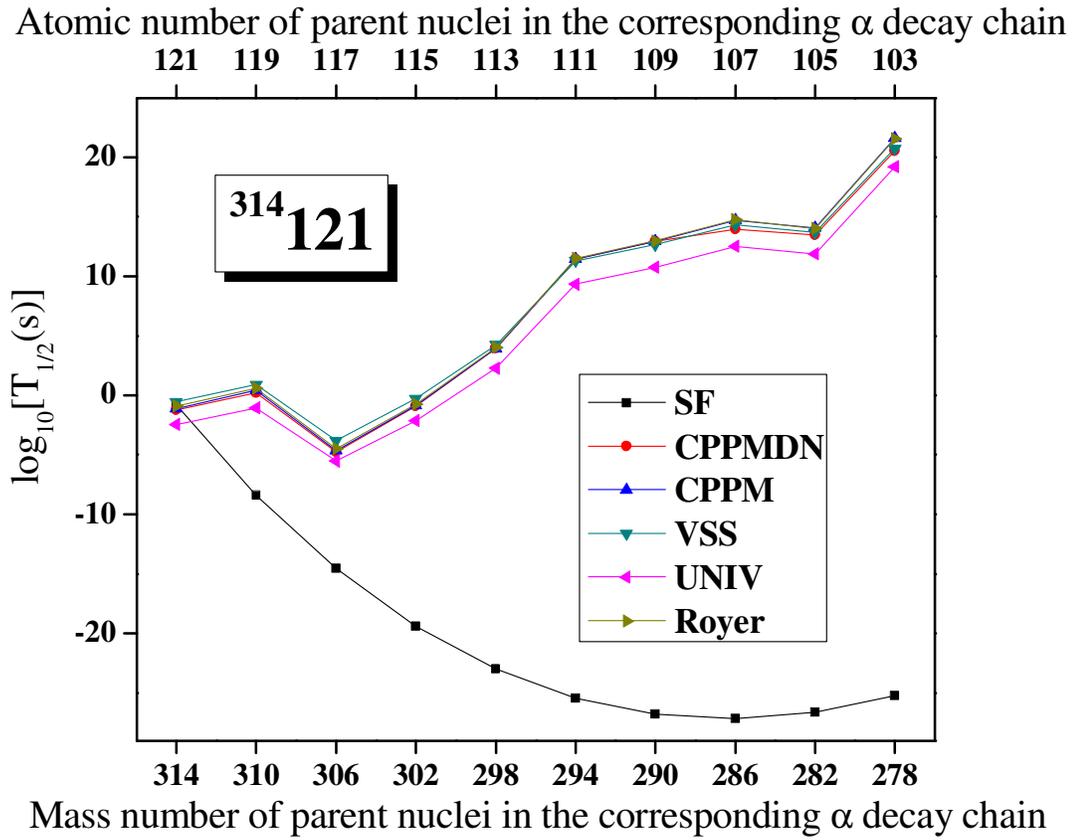

Fig 6: The comparison of the calculated alpha decay half-lives with the spontaneous fission half-lives for the isotope $^{314}121$ and its decay products.

Table I: Predictions on the mode of decay of $^{309-314}$121 superheavy nuclei and their decay products by comparing the alpha half lives and the corresponding spontaneous fission half lives. $T_{SF}^{av}$ is calculated using Ref [44].

| Parent Nuclei | $Q_\alpha$(Cal) MeV | $T_{SF}^{av}$ (S) | T$_{1/2}$(s) | | | | | Mode of Decay |
| --- | --- | --- | --- | --- | --- | --- | --- | --- |
| | | | CPPMDN | CPPM | VSS | UNIV | Royer | |
| $^{309}$121 | 13.118 | 5.574x10$^9$ | 1.376x10$^{-5}$ | 1.376x10$^{-5}$ | 3.586x10$^{-5}$ | 1.847x10$^{-6}$ | 6.844x10$^{-6}$ | α |
| $^{305}$119 | 13.916 | 1.585x10$^2$ | 8.145x10$^{-8}$ | 8.145x10$^{-8}$ | 3.315x10$^{-7}$ | 2.418x10$^{-8}$ | 5.955x10$^{-8}$ | α |
| $^{301}$117 | 11.994 | 1.197x10$^{-4}$ | 3.232x10$^{-4}$ | 3.232x10$^{-4}$ | 7.549x10$^{-4}$ | 3.284x10$^{-5}$ | 1.305x10$^{-4}$ | SF |
| $^{310}$121 | 11.928 | 2.787x10$^9$ | 5.204x10$^{-3}$ | 8.072x10$^{-3}$ | 3.060x10$^{-2}$ | 4.911x10$^{-4}$ | 1.345x10$^{-2}$ | α |
| $^{306}$119 | 13.986 | 7.927x10$^1$ | 5.699x10$^{-8}$ | 5.740x10$^{-8}$ | 5.438x10$^{-7}$ | 1.788x10$^{-8}$ | 1.009x10$^{-7}$ | α |
| $^{302}$117 | 12.804 | 5.985x10$^{-5}$ | 4.305x10$^{-6}$ | 4.305x10$^{-6}$ | 3.121x10$^{-5}$ | 7.579x10$^{-7}$ | 6.707x10$^{-6}$ | α |
| $^{298}$115 | 9.653 | 9.770x10$^{-10}$ | 1.937x10$^2$ | 2.333x10$^2$ | 5.155x10$^2$ | 6.391x10$^0$ | 3.090x10$^2$ | SF |
| $^{311}$121 | 11.878 | 9.784x10$^4$ | 9.109x10$^{-3}$ | 1.042x10$^{-2}$ | 1.824x10$^{-2}$ | 6.127x10$^{-4}$ | 3.218x10$^{-3}$ | α |
| $^{307}$119 | 13.446 | 2.616x10$^{-3}$ | 6.752x10$^{-7}$ | 6.812x10$^{-7}$ | 2.509x10$^{-6}$ | 1.456x10$^{-7}$ | 4.165x10$^{-7}$ | α |
| $^{303}$117 | 12.784 | 1.856x10$^{-9}$ | 4.610x10$^{-6}$ | 4.610x10$^{-6}$ | 1.559x10$^{-5}$ | 7.987x10$^{-7}$ | 2.467x10$^{-6}$ | SF |
| $^{312}$121 | 11.798 | 4.892x10$^4$ | 1.794x10$^{-2}$ | 1.607x10$^{-2}$ | 6.195x10$^{-2}$ | 8.952x10$^{-4}$ | 2.637x10$^{-2}$ | α |
| $^{308}$119 | 12.216 | 1.308x10$^{-3}$ | 2.786x10$^{-4}$ | 3.572x10$^{-4}$ | 1.893x10$^{-3}$ | 3.302x10$^{-5}$ | 5.620x10$^{-4}$ | α |
| $^{304}$117 | 12.874 | 9.282x10$^{-10}$ | 2.839x10$^{-6}$ | 2.839x10$^{-6}$ | 2.254x10$^{-5}$ | 5.219x10$^{-7}$ | 4.377x10$^{-6}$ | SF |
| $^{313}$121 | 11.708 | 3.326x10$^{-1}$ | 2.506x10$^{-2}$ | 2.644x10$^{-2}$ | 4.625x10$^{-2}$ | 1.385x10$^{-3}$ | 7.533x10$^{-3}$ | α |
| $^{309}$119 | 11.756 | 8.362x10$^{-9}$ | 4.635x10$^{-3}$ | 4.635x10$^{-3}$ | 9.645x10$^{-3}$ | 3.147x10$^{-4}$ | 1.491x10$^{-3}$ | SF |
| $^{314}$121 | 11.518 | 1.663x10$^{-1}$ | 5.876x10$^{-2}$ | 7.999x10$^{-2}$ | 2.947x10$^{-1}$ | 3.696x10$^{-3}$ | 1.287x10$^{-1}$ | α |
| $^{310}$119 | 10.726 | 4.181x10$^{-9}$ | 1.667x10$^0$ | 2.671x10$^0$ | 8.232x10$^0$ | 9.240x10$^{-2}$ | 3.968x10$^0$ | SF |